\documentstyle[preprint,epsfig,aps]{revtex}

\begin{document}

\title{Modelling Horizontal Air Showers Induced by Cosmic Rays}

\author{M. Ave, R.A. V\'azquez, and E. Zas}

\address{Departamento de F\'\i sica de Part\'\i culas, Universidade 
de Santiago E-15706 Santiago de Compostela, Spain}

\maketitle

\begin{abstract}
We present a framework for the study of muon density patterns on 
the ground due to showers produced in the atmosphere by cosmic rays 
incident at high zenith angles. As a checking procedure predictions 
of a model based on such a framework are compared to simulation 
results. 
\end{abstract}

{\bf PACS number(s):} 96.40.Pq; 96.40.Tv

{\bf Keywords:}

\section{Introduction}

The old idea to detect neutrinos through Horizontal Air Showers (HAS) 
\cite{Berezinskii} has been recently brought to attention with the 
calculation of the acceptance of the Auger Observatories for the 
detection of high energy neutrinos \cite{Capelle_1}. At large zenith 
angles, cosmic rays (whether they are protons, heavier nuclei, or 
even photons) develop ordinary showers in the top layers of the 
atmosphere in a very similar fashion to the well understood vertical 
showers. Their electromagnetic component is however almost 
completely absorbed by the greatly enhanced atmospheric slant 
depth and prevented from reaching ground level. By contrast, high 
energy neutrinos may also induce Horizontal Air Showers (HAS) deeper 
into the atmosphere and hence resemble vertical air showers in their 
particle content at ground level, provided they interact 
sufficiently close to an air shower array. 

The main background to neutrino induced HAS of very high energy 
(in the EeV range and above) is expected to be due to the remaining 
muon component of the cosmic ray showers, after practically all the electromagnetic component is absorbed. 
A preliminary study of the air shower data at zenith angles above $60^o$ 
from the Haverah Park array has shown that the observed 
rate is consistent to that expected from the ultra high 
energy ($E > 1$~EeV) cosmic rays arriving at large zenith 
angles \cite{Ave_1}. These {\sl muon showers} that 
penetrate the whole atmospheric slant depth to reach ground level are 
the object of this study which is intended to be detector independent. 
As the neutrino induced HAS rate that could be expected is quite low, 
zenith angle misreconstructions of vertical showers can also constitute 
a serious potential background to neutrino detection. 
However the relative importance of this background is very dependent
on technical details of the experimental setup and, in particular, on
the angular resolution actually achieved. Thus these matters cannot be
further addressed in this article.

The interest in HAS induced by ordinary cosmic rays is 
not just limited to understanding the background of neutrino induced showers. 
The measurement of high zenith angle showers would enhance the aperture 
of existing air shower array data and it would increase the data on cosmic 
ray arrival directions to previously inaccessible directions in galactic 
coordinates \cite{Ave_1}. 
Besides these obvious advantages, high zenith angle cosmic ray showers 
are unique because they select the muon component of the showers and they 
naturally probe a region of phase space which is quite inaccessible 
from vertical air shower measurements: the shower core. 

The cosmic ray induced HAS are different mainly because they consist of 
muons which are produced far away from ground level. 
The particle density profiles for HAS induced by protons 
or heavy nuclei display complex muon patterns at the ground which 
result from the long path lengths travelled by the muons in the 
presence of the Earth's magnetic field \cite{Hillas,Andrews}. 
These complex patterns are difficult to analyse \cite{Dedenko} because 
the conventional approaches used for interpretation of low zenith 
angle showers ($< 60^\circ$) are usually based on the approximate 
circular symmetry of the density profiles. The analysis of HAS produced 
by cosmic rays requires a radically different approach. 

Here we present a qualitative explanation of the ground muon patterns 
as well as analytical expressions for the muon density profile based on 
simple ideas. Our approach agrees remarkably well with simulation results 
for a wide range of distances to the core and zenith angles between 
$60^\circ$ and $90^\circ$. 
Besides its convenience for data analysis which would otherwise depend on 
lengthy and cumbersome simulation results, the model gives a lot of insight 
into HAS and their complementarity to vertical showers 
and can be used for calculating the expected HAS rates 
for future detectors such as the Auger Observatories \cite{Ave_prep}. 
We also briefly discuss a simple method involving rotations for approximately 
generating muon ground density profiles for any azimuthal incidence using 
a single shower of the same zenith angle. 

The article is organized as follows. 
In section \ref{toy} we present a toy model which describes the qualitative 
behavior of the ground muon patterns. 
In section \ref{geomag} we discuss quite generally the implications 
of geomagnetic fields and the geometry implied. 
In section \ref{geotoy} we extend the toy model to account for 
the magnetic field. 
In section \ref{param} we give a detailed prescription to calculate muon densities
for any azimuthal and zenith angle. In section \ref{asymmetries} we 
discussed the asymmetries that arise at ground level and introduce 
a number of ways to account for them. In section \ref{results} 
we compare the analytical densities obtained to those obtained with 
Monte Carlo simulations 
and finally in section \ref{summary} we summarize the achievements, 
briefly enumerating possible applications. 

\section{Generalities: A Toy Model}
\label{toy}

The number of electrons, positrons and photons, the electromagnetic 
component, in an air shower induced by a 10$^{19}$ eV proton rises 
as the shower penetrates higher depths to reach a maximum after 
which it rapidly becomes absorbed by low energy processes and the 
photoelectric effect. This component 
behaves much like shower size (number of electrons and positrons) 
reaching its maximum typically at $X_{m} = 750$~g~cm$^{-2}$. 
Below this depth it starts to become exponentially suppressed. 
The atmosphere is $\sim$ 1000 g~cm$^{-2}$ deep for vertical 
showers doubling to $\sim$ 2000 g~cm$^{-2}$ for 60$^\circ$ and 
becoming over 30 times deeper for completely HAS at 
sea level so that the electromagnetic component 
from $\pi^0$ decay can be safely neglected at large zenith angles. 

For inclined showers the main sources of electrons and photons at 
ground level are in fact the long range muons themselves, 
mainly through decay, but also through bremsstrahlung and pair 
production. 
The electron number density follows closely that of muons at all 
distances from shower axis and, as they are produced in small 
electromagnetic subshowers initiated by the decay electrons, their 
energy spectrum is very much like that expected in a vertical shower. 
A very similar argument applies to photons so that we can state 
that the total electromagnetic energy of an inclined shower is 
much less than the energy of the muonic component, unlike in vertical 
showers. 

As the zenith angle rises, the average length traversed by 
the shower muons from their production point to the ground 
increases from $\sim 10~$km for vertical showers to 
$ \sim 300$~km 
for nearly horizontal ones (see table~\ref{alturas}) . 
As a result only the muons produced with sufficiently high energy 
survive to ground level and this is crucial for HAS. 
Typically the energy loss ranges from a few GeV 
for the lowest energy muons in vertical showers to well over 100~GeV 
for a completely horizontal shower. As a result the average energy 
of the muons at ground level varies by close to two orders of 
magnitude between vertical and horizontal showers, as shown in 
table~\ref{alturas}.


We now introduce a very simple toy model that reproduces some of the 
qualitative features of the muon density patterns on the ground. Later we will 
build on this simple model to obtain parameterizations of muon densities at 
ground level. 
Let us consider a shower of zenith angle $\theta$ and assume that 
there is no magnetic field. At ground level we calculate muon number 
densities 
in a plane perpendicular to the shower axis: {\sl the transverse plane}. 
In this plane the description of the shower becomes considerably 
simpler because it accurately respects the approximately cylindrical symmetry 
expected, except for minor effects due to changes in air density 
with height. This plane avoids the distortion of 
the muon densities resulting from the projection onto the Earth surface. 
Moreover, 
when the magnetic field is fully taken into account and the cylindrical 
symmetry is broken, as the azimuthal angle of the shower is rotated 
the muon density contours just rotate in the transverse plane 
remaining otherwise practically unchanged as will be shown below. 

We consider a highly relativistic muon of energy $E \simeq c p$, 
and transverse momentum $p_{\perp}$ that travels a
distance $d$ to reach the ground. We assume this muon will only be 
deviated due to the $p_{\perp}$ inherited from its parent pion (kaon) to 
arrive at distance $r$ to shower axis in the transverse plane. 
This is equivalent to neglecting multiple scattering and deviations 
of the pions themselves from shower axis. 
In this case $r$ is simply given by:
\begin{equation}
r=  \frac{p_{\perp}}{p}~d \simeq \frac{c p_{\perp} }{E}~d.
\label{rE}
\end{equation}
This relation means that the muons of a given shower have distributions 
in $E$, $d$ and $p_{\perp}$ which are not independent. 

For a given zenith $\theta$ the toy model is obtained taking an input 
energy distribution for the shower muons and fixing the values of $d$ 
and $p_{\perp}$. 
The production distances $d$ of the shower muons reaching ground level 
have a distribution fixed by shower development with an average 
$< d >$ that rises as the zenith angle $\theta$ increases. 
As the development of inclined showers takes place in the upper layers 
of the atmosphere and spans only 
a small fraction of the atmosphere, the width of the $d$ distribution is 
small compared to $< d >$. This means that taking $d$ to be a function of 
$\theta$ and ignoring its distribution can be a fairly good approximation. 
On the other hand the muon $p_{\perp}$ results from multiple pion interactions in the shower development and has a relatively broad distribution. This is 
one of the reasons that prevents the toy model from being valid at 
quantitative level as will be shown below. 

\begin{figure}[hbt]
\centering
\mbox{\epsfig{figure=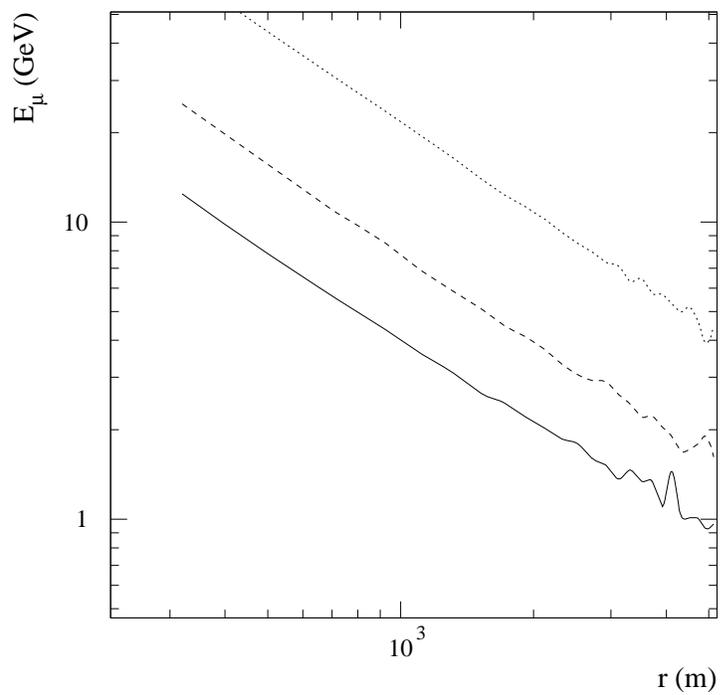,width=11.0cm}}
\caption{Average muon energy as a function of distance to shower core. 
Results obtained with simulation using Aires for $10^{19}~$eV proton 
showers of zenith angle 60$^\circ$, 70$^\circ$, and 80$^\circ$ from 
bottom to top. }
\label{fig1}
\end{figure}

The virtue of this toy model is that it implies an approximately 
correct correlation between the muon energy, $E$, and its deviation 
from shower axis, $r$, through Eq.~(\ref{rE}). Fig.~\ref{fig1} shows 
the correlation between the average muon energy and $r$ as obtained 
in the simulation of one hundred $10^{19}~$eV proton showers 
without magnetic field effects. \footnote{This and all the simulations 
in this work, unless otherwise stated, have been generated with 
AIRES (verison 1.4.2) \cite{Aires} and SIBYLL 
(version 1.6) \cite{Sibyll} for hadronic interactions.} 
Such a relation means that at 
ground level the muon differential energy spectrum, $\phi[E]=dN/dE$, 
and the number density, $\rho_{\mu}(r)$, are related through:  
\begin{equation}
\rho_{\mu}(r)=  -\frac{d E}{dr} \frac{1}{2 \pi r} \phi \left[E(r)\right] = 
         \frac{ c p_{\perp} d}{2 \pi r^3} \; \phi \left[E(r)\right].
\end{equation}
If for instance the energy spectrum of the shower muons is assumed to be 
$\phi(E) = A E^{-\gamma}$ the above relation implies: 
\begin{equation}
\rho(r)= A ~ \frac{(c p_{\perp} d)^{1-\gamma}}{2 \pi } ~ r^{-3 +\gamma}.
\end{equation}

\begin{figure}[hbt]
\centering
\mbox{\epsfig{figure=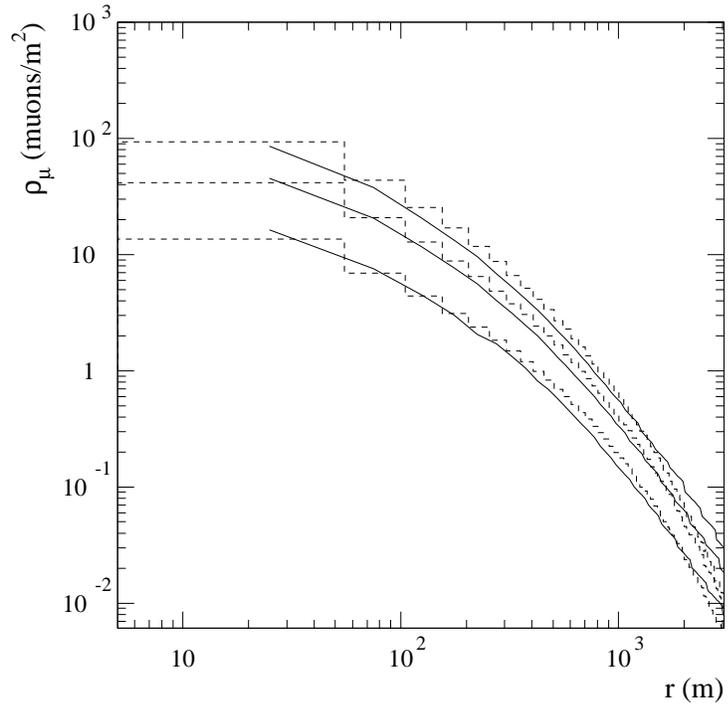,width=11.0cm}}
\caption{Lateral distribution of muons as obtained in the toy model 
(continuous lines) and in the simulation (dashed lines). Figures 
correspond from top to bottom to zenith angles 60$^\circ$, 
70$^\circ$, and 80$^\circ$.}
\label{fig2}
\end{figure}

\begin{figure}[hbt]
\centering
\mbox{\epsfig{figure=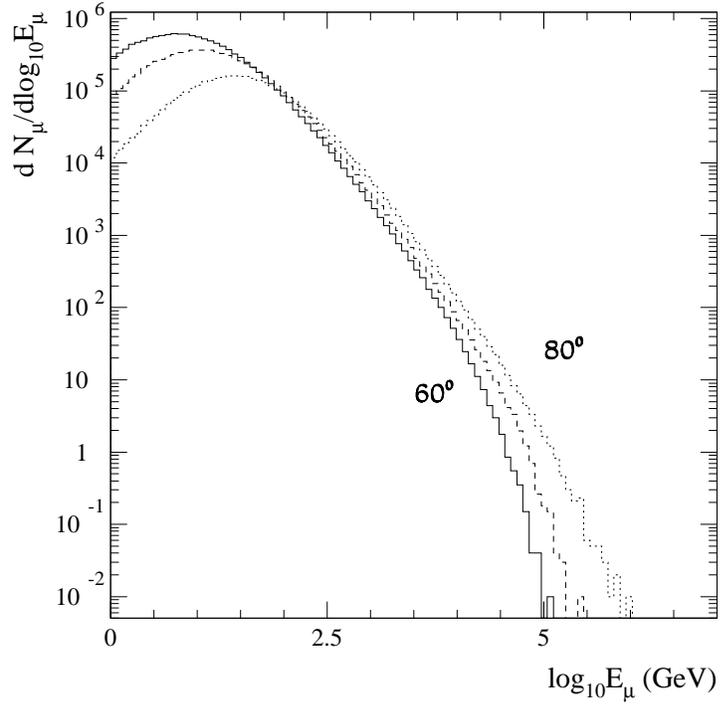,width=11.0cm}}
\caption{Energy spectrum of muons at ground level for $10^{19}~$~eV 
proton showers simulated with AIRES, for zenith angles 
60$^\circ$ (full line), 70$^\circ$ (dashed line), 
and 80$^\circ$ (dotted line).}
\label{fig3}
\end{figure}

In the absence of a geomagnetic field the toy model can be adjusted to 
roughly reproduce Monte Carlo results. One can use a muon energy spectrum 
$\phi[E]=dN/dE$ and $d$ as obtained in simulations and only needs to adjust 
a single parameter namely $p_{\perp}$. 
We have done this exercise to obtain muon densities which 
are compared to AIRES simulations at different zenith angles. 
As an example Fig.~\ref{fig2} compares the results of simulation for 
three different zenith angles to density profile functions 
obtained with the toy model using the muon energy distributions 
from the simulations which are shown in Fig.~\ref{fig3}. 
It is remarkable that such a crude model reproduces the muon lateral 
distributions to better than $30 \%$ in the interval 20~m $< r <$ 2~km 
with a single $p_{\perp}$ value of $200~$MeV for all zenith 
angles considered. The results are 
nevertheless only academic since in practice the ground density 
profiles are severely modified by the Earth's magnetic field.

\section{Generalities: Geomagnetic deviations}
\label{geomag}

\begin{figure}[hbt]
\centering
\mbox{\epsfig{figure=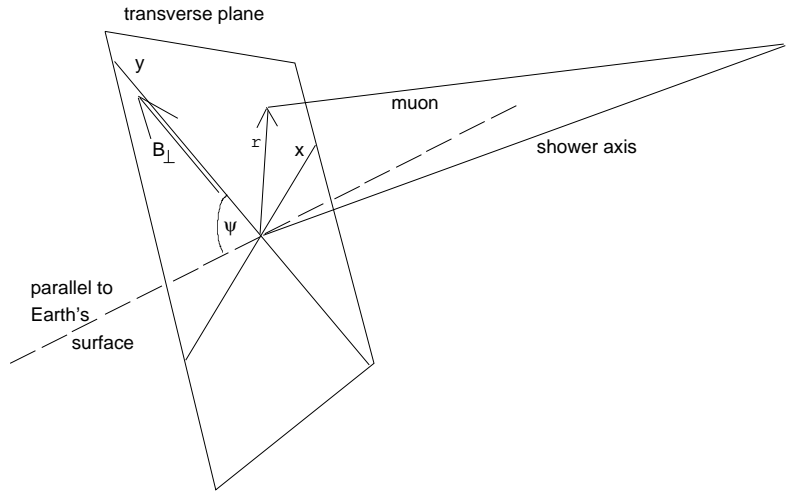,width=11.0cm}}
\caption{The plane perpendicular to the shower axis or 
{\sl transverse plane} and some useful geometrical definitions 
for Horizontal Air Showers. The $y$ axis is chosen along the direction of 
$B_{\perp}$ which is the projection of the magnetic field onto the 
transverse plane. $\psi$ is the angle subtended by the $y$ axis and 
the intersection of the transverse and the ground planes.}
\label{geometry}
\end{figure}

To study the effect of the Earth's magnetic field we choose to decompose 
the field $\vec B$ in two components 
$\vec B_{\parallel}$ and $\vec B_{\perp}$, respectively parallel and 
perpendicular to shower axis. It is worth studying with a 
little detail the resulting geometry in the transverse plane 
since much of the ground pattern can be qualitatively understood 
in this way. We define two coordinate axes ($x,y$) 
in this plane, such that $y$ is parallel to $\vec B_{\perp}$ 
as shown in Fig.~\ref{geometry}. In the transverse plane 
we define $\psi$ as the angle between the $y$ axis
(or equivalently $\vec B_{\perp}$) and the direction parallel 
to the Earth's surface.

A muon travelling a distance $d$ from the production site to the ground 
will bend its trajectory in a direction perpendicular to its velocity, 
$\vec v$ and to $\vec B_{\perp}'$. Here $\vec B_{\perp}'$ is the 
component of $\vec B$ perpendicular to $\vec v$. If the muons are 
almost parallel to shower axis $\vec B_{\perp}'$ largely coincides 
with $\vec B_{\perp}$ and the transverse plane is perpendicular to the 
muon trajectory in this approximation. 
The parallel component of the Earth's magnetic field 
$\vec B_{\parallel}$ can be neglected and neglecting other muon 
interactions and energy loss, the helicoidal trajectory 
is well approximated by an arc of a circle of radius $R$ in the plane 
containing shower axis and $\vec B_{\perp}$. 

For a complete description 
of the deviations we only need $\vec B_{\perp}$ or, equivalently, its 
intensity $B_{\perp}$ and the angle $\psi$. 
The circular symmetry of the shower in the absence of magnetic 
fields is distorted depending on the projected 
magnetic field $B_{\perp}$, the distance 
traversed by the muons, and their energy distribution. At low 
zenith the path length travelled is small and the circular patterns 
are hardly modified because the deviations are small. 
As the zenith rises the pattern turns first 
into an ellipsoid which eventually becomes two lobes at 
each side of $B_{\perp}$ corresponding to the negative and positive 
charge muons deviating in opposite directions. 

\begin{figure}[hbt]
\centering
\mbox{\epsfig{figure=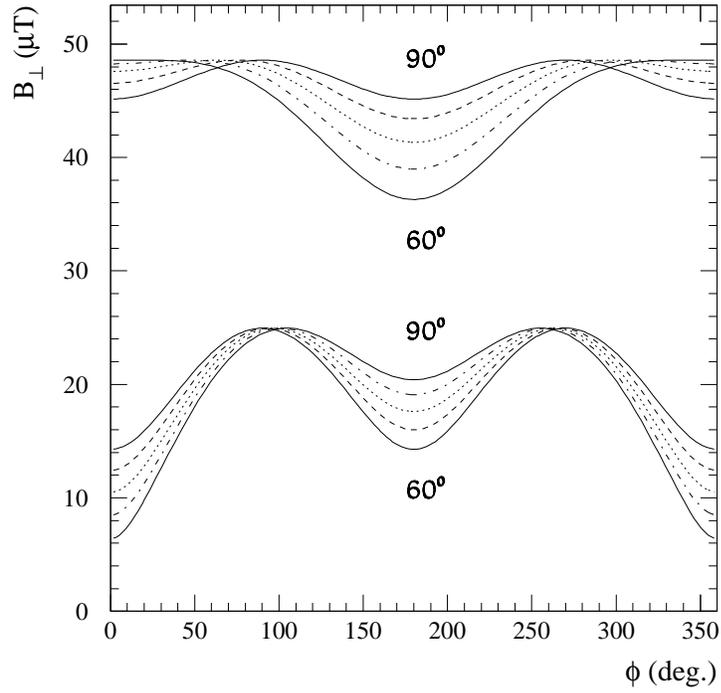,width=11.0cm}}
\caption{Magnetic field projection onto the transverse plane 
as a function of azimuthal angle for the Haverah Park 
location (top lines) and Pampa Amarilla (bottom lines). Lines are 
for zenith angles 60$^\circ$, 70$^\circ$, 80$^\circ$, 87$^\circ$, 
and 90$^\circ$.}
\label{Bmod}
\end{figure}

\begin{figure}[hbt]
\centering
\mbox{\epsfig{figure=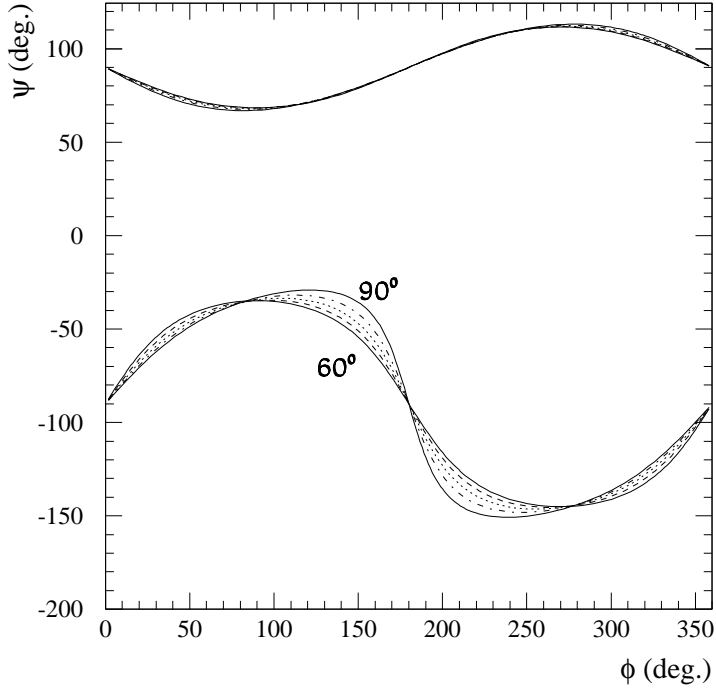,width=11.0cm}}
\caption{Angle $\psi$ in the transverse plane (between 
$\vec B_{\perp}$ and the line parallel to the surface of the Earth, 
see Fig.~4) 
as a function of azimuthal angle for the Haverah Park location 
(top lines) and Pampa Amarilla (bottom lines).
Lines are for zenith angle 60$^\circ$, 70$^\circ$, 80$^\circ$,
87$^\circ$, and 90$^\circ$.} 
\label{Bpsi}
\end{figure}

For a fixed zenith angle the density pattern in the transverse plane  
depends only on the intensity of the magnetic field. The pattern 
is oriented according to the direction of $\vec B_{\perp}$. 
Both the magnitude and direction of $\vec B_{\perp}$ depend on 
the azimuthal angle of the shower $\phi$. 
In this article azimuthal angles are measured counterclockwise 
with respect to the magnetic North. 
Figs.~\ref{Bmod} and \ref{Bpsi} respectively 
show how both these 
quantities vary as the azimuthal angle takes all possible values 
at two different locations, corresponding to Haverah Park 
\cite{Haverah} and Pampa Amarilla (the southern hemisphere Auger 
Observatory) \cite{Auger} for illustrative purposes. On the basis 
of these two graphs alone and the discussion that follows it is 
straightforward to conclude that the 
muon density patterns in these sites will be completely different. 
A detailed study of HAS in Pampa Amarilla will be 
published elsewhere. 

As the shower azimuthal angle varies at the Haverah Park site, 
$B_{\perp}$ oscillates in magnitude about a central value. The oscillation 
amplitude decreases as the zenith angle rises being $15\%$ ($8\%$) 
of the central value for $70^\circ$ ($80^\circ$) zenith. 
It is remarkable that for this location once the zenith 
angle $\theta$ is fixed at a high value, $B_{\perp}$ is 
not too sensitive to the azimuthal direction of the shower $\phi$. 
On the other hand as the zenith angle decreases the length transversed 
by the muons $d$ becomes much smaller and the magnetic field is 
less effective in modifying the ground density profiles. While for 
low zenith 
the magnetic effect is small and there is little need for considering 
it, for high zenith $B_{\perp}$ can be 
regarded as approximately independent of azimuth. 

This means that we can make use of the fact that $B_{\perp}$ 
is approximately azimuth independent to quickly generate HAS 
from a single simulation. 
If we fix the zenith angle to a high value, we make the approximation 
that $B_{\perp}$ is constant and let the azimuth angle vary, the 
variations of the shower densities at ground level are only due to 
the corresponding changes in the angle $\psi$. In this approximation 
different azimuths are equivalent to a pattern rotation in the transverse 
plane. This approximation has important practical implications 
as it means that, given the zenith angle, a single shower can be used 
to generate muon density profiles on the ground for all azimuthal angles 
with the appropriate rotations of $\vec B_\perp$ and projections onto the 
plane corresponding to the Earth's surface. This alone represents an 
important tool that can help the analysis of high energy HAS and 
has already been successfully used in the analysis of Haverah Park data 
presented in \cite{Ave_1}. 

We have tested this economical means for generating HAS in a very 
simple fashion. We have generated approximate showers of predetermined 
azimuths from the results of a single shower. The ground 
level muon number density contours obtained in this way have been 
compared with the results of a direct simulations for the corresponding 
rotated azimuths. For the Haverah Park site as an example we have obtained 
approximate proton showers at $\phi=90^\circ$ and $180^\circ$ from the 
Monte Carlo 
results of a shower at $\phi=0^\circ$ for three different zenith angles: 
$\theta=60^\circ,70^\circ,80^\circ$. The rotated showers agree rather 
well with simulation results. Indeed fluctuations from shower to 
shower and the uncertainties introduced by thinning are much larger than 
the error made in the constant $B_{\perp}$ approximation. 
At extreme zeniths ($\sim 87^{\circ}$) 
however the separated lobes on the ground plane obtained with simulations 
are slightly asymmetric. Clearly rotation in the transverse plane and 
simple projection onto the ground cannot generate these asymmetries. 
These asymmetries arise because at high zeniths most of the muon 
deviation is due to the magnetic field and linear projection becomes 
inadequate. They can be taken into account as will be further discussed 
in section \ref{asymmetries}. 

\section{Geomagnetic deviations in the Toy model}
\label{geotoy}

\begin{figure}[hbt]
\centering
\mbox{\epsfig{figure=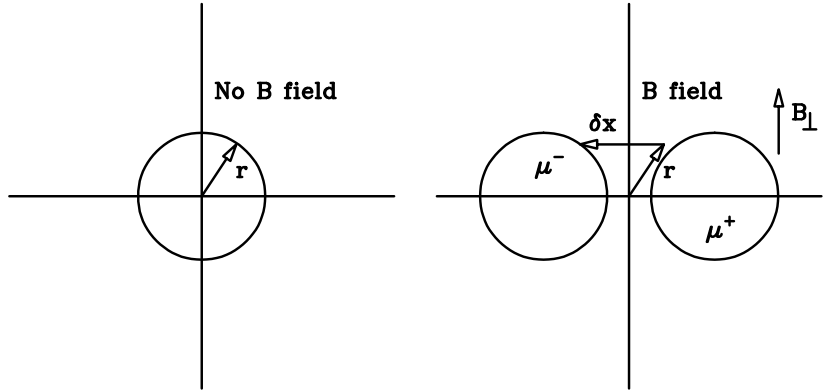,width=11.0cm}}
\caption{Toy model composition of deviations due to transverse 
momentum and to the geomagnetic field. The second graph 
illustrates how the positive and negative muons that would be a distance 
$r$ away from shower axis in the absence of magnetic field, are split 
and deviated an amount $\delta x$ into two opposite directions 
along the 
$x$ axis (perpendicular to $\vec B_{\perp}$).}
\label{symmetry}
\end{figure}

The distinction between the two parameters $p_{\perp}$ and $d$ made in 
section \ref{toy} is important for considering geomagnetic field effects 
separately. We quantify the magnetic deviation through $\delta x$,  
the distance between the intersections of the two muon trajectories 
with and without magnetic effects and the transverse plane as shown in 
Fig.~\ref{symmetry}.
Note that the coordinates $x,y$ are conveniently chosen in the 
perpendicular and parallel directions to $\vec B_{\perp}$. 

The geomagnetic deviation of the muons is easily shown to be:
\begin{equation}
{\delta x}= R \left[1- \sqrt{1-\left(\frac{d}{R}\right)^2} \right] 
\simeq \frac {d^2} {2 R} = \frac{e B_{\perp} d^2}{2p},
\label{deltax}
\end{equation}
where $e$ is the electron charge, $R$ is the radius of curvature, 
and $p$ is the muon momentum and we have expanded brackets to first 
order in $d/R$. For $\delta x$ less than 
2~km we expect it to be valid at the $5 \%$ level at zenith $60^{\circ}$ 
because the typical production distance $< d >$ is 16~km. At higher 
zenith the approximation is excellent. 
When we combine the above relation with Eq.~\ref{rE} we obtain the 
following expression:
\begin{equation}
\delta x = \frac {e B_\perp d^2} {2p} = 
\frac{0.15 B_\perp d}{p_{\perp}} \; \bar r = \alpha \; \bar r,
\label{alpha}
\end{equation}
where in the last equation $B$ is to be expressed in Tesla, $d$ in m and 
$p_{\perp}$ in GeV. 

Physically $\bar r$ (given by Eq.~\ref{rE}) corresponds to the muon 
deviation in the transverse plane {\sl in the absence of magnetic field}. 
When $\bar r$ and $\delta x$ are small compared to $d$ we can add 
them as two dimensional vectors in this plane. 
We can gain some insight in this relation considering that 
all muons (that for $\vec B=0$ would intersect the transverse plane 
at a distance $\bar r$ to the shower axis) are further deviated 
by $\vec B \neq 0$ a distance proportional to $\bar r$ in the 
direction perpendicular to $\vec B_{\perp}$ as schematically shown in 
Fig.~\ref{symmetry}. 
The $\mu^+$ and $\mu^-$ deviate in opposite senses so that 
the projection of the geomagnetic field in the transverse plane 
introduces a kind of mirror symmetry axis. We expect that 
the symmetry will not be exact being modulated by the expected excess of 
$\mu^+$ respect to $\mu^-$ in atmospheric showers \cite{Frazer}. 

The dimensionless parameter $\alpha$ measures the ratio of the 
two described components of the displacement in the transverse plane, 
that due to the transverse momentum and that due to the magnetic field. 
Technically we can obtain the muon number density in the 
transverse plane in this model making the transformation:  
\begin{equation}
\begin{array}{ll}
x= &  \bar x + \alpha \sqrt{\bar x^2+\bar y^2}, \\
y= & \bar y. \\
\end{array}
\label{transform}
\end{equation}
Here the barred (unbarred) coordinates represent the position of the 
muon in the transverse plane in the absence (presence) of the Earth's 
magnetic field. The muon number density is then: 
\begin{equation}
\rho(x,y) = \bar \rho \left( \bar x(x,y),\bar y(x,y) \right) \; 
\left[ \frac{\partial (\bar x \bar y)} {\partial (x y)} \right],
\label{rhotoy}
\end{equation}
where $\bar \rho(\bar x, \bar y)$ is the density at distance 
$\bar r=\sqrt{\bar x^2+\bar y^2}$ 
in the case $\vec B=0$ and the last factor in brackets is the Jacobian 
of the transformation. 
We have checked that the above equations reproduce the main 
features observed in the muon density profiles in the transverse 
plane with magnetic fields. The muon number densities at ground 
level can be obtained after an appropriate projection onto the 
plane of the Earth's surface. 

The value of $\alpha$ is a measure of the importance of the 
magnetic field. For instance for vertical showers we can plausibly 
set the distance to production to $d \sim 4 $~km, $ B_{\perp} 
\sim 10~\mu$T, and $p_{\perp} \sim 0.3 $~GeV, which gives $\alpha 
\sim 0.02$; {\it i.e.} $\alpha$ is very small and the effect of the
magnetic field is not very important for vertical showers. For 
horizontal showers, however, $d$ can be as large as $300$ km and 
$B_{\perp} \simeq 40~\mu$T, making $\alpha \gg 1$. For large values 
of $\alpha$ we expect that the density profiles obtained with 
$\vec B=0$ are strongly modified. 

\begin{figure}[hbt]
\centering
\mbox{\epsfig{figure=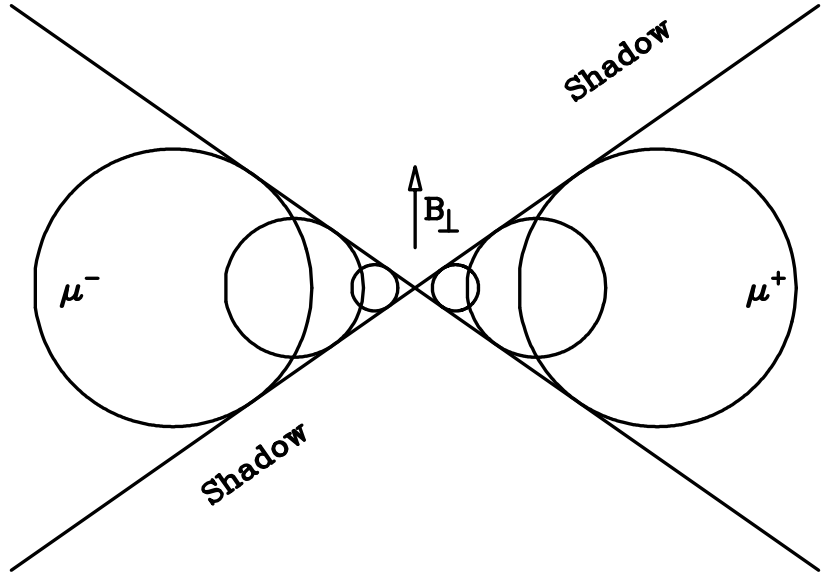,width=11.0cm}}
\caption{Toy model magnetic deviations in the transverse plane 
associated to muons of 
different energies (and different radii), illustrating the 
origin of the {\sl shadow} regions in which 
no muons are expected. The geometrical construction is identical 
to the two dimensional plot typically used to explain Cherenkov 
radiation.}
\label{Cherenkov}
\end{figure}

\begin{figure}[hbt]
\centering
\mbox{\epsfig{figure=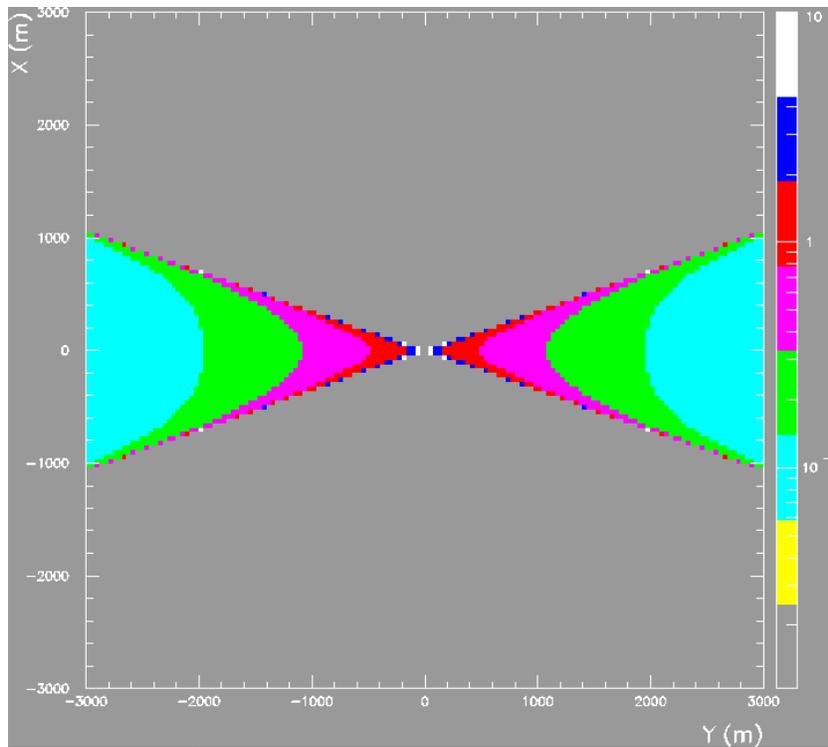,width=11.0cm}}
\caption{Toy model results for the muon number densities in the 
transverse plane with shadow regions as described in the text.}
\label{BTOY}
\end{figure}

It can be shown that if $\alpha > 1$ all $\mu^+$ and $\mu^-$ 
are confined to a region in the transverse plane limited by two 
straight lines as illustrated in Fig.~\ref{BTOY}. The figure has 
been obtained with Eq.~\ref{rhotoy}, $\alpha=3$ and using a standard 
parameterization for the lateral muon distribution: 
$\rho(r) \propto r^{-0.5} (1+ r/a)^{-5}$ with $a=1.5~$km. 
This result is qualitatively very 
important because it implies that for large zenith angles we expect 
{\sl void regions} or {\sl shadows} in which no muons are expected. 
The angle subtended by these voids is dependent on the parameter 
$\alpha$, physically on the zenith angle. The effect can be easily 
explained with a geometrical construction matching the textbook 
graph for explaining the Cherenkov effect, see Fig.~\ref{Cherenkov}. 
The existence of these shadows is indeed the case as observed in 
simulations (see next section) and can be used for distinguishing 
between HAS produced by protons or nuclei high in the atmosphere 
and those produced by deeply penetrating neutrinos. These results 
are only valid at a qualitative level and hence the actual 
function used for $\rho(r)$ is not relevant. Further steps are needed to 
transform the toy model into a valid parameterization as shown below. 

\section{Muon Number Densities in HAS}
\label{param}

One of the reasons that has difficulted the study of HAS in particle array 
experiments is their intricate density profiles and their 
azimuthal dependence. Experiments with good sensitivity to particles 
incident at high zenith angles such as the Haverah Park Array did not 
have data from many stations that allowed the study of these effects 
from the experimental data. From the discussion of the 
previous section it is clear that the expected asymmetries due to the 
geomagnetic 
field depend on the value 
of the projection of $\vec B$ onto the transverse plane. 
This projection obviously differs depending on the location of the experiment. 
For the purpose of checking the validity of the parameterizations 
that we present here, we have selected as an example the field corresponding 
to the Haverah Park array site, which has a value of 
$48.5 \; \mu$T, declination of $-7.61^ \circ$ and inclination of 
$68.4^\circ$. 

Unfortunately the toy model presented in the previous section 
is not good enough for quantitative 
purposes. The actual situation is more complicated even when $\vec B=0$ 
because the transverse position of the muon, $\bar r$, is affected 
by both multiple scattering and the transverse position of the parent pions. 
Nevertheless the simple relation provided by Eq.~\ref{rE} manifests a 
true correlation between $\bar r$ and $E$ which can be easily checked 
through simulation. 
Fig.~\ref{fig1} shows how the {\sl average} muon energy, $<E>$,  
is clearly inversely related to distance from the shower 
axis $\bar r$ as obtained in AIRES simulated 
proton showers at zenith angles $60^\circ$, $70^\circ$ and $80^\circ$. 
To a very good approximation 
$\bar r \propto <E>^{-\beta}$ where $\beta \sim 1.1 $. 
We will now show how a relatively simple extension of the toy model 
converts it into a quantitatively valid parameterization of the muon 
component in HAS induced by cosmic rays. 

\begin{figure}[hbt]
\centering
\mbox{\epsfig{figure=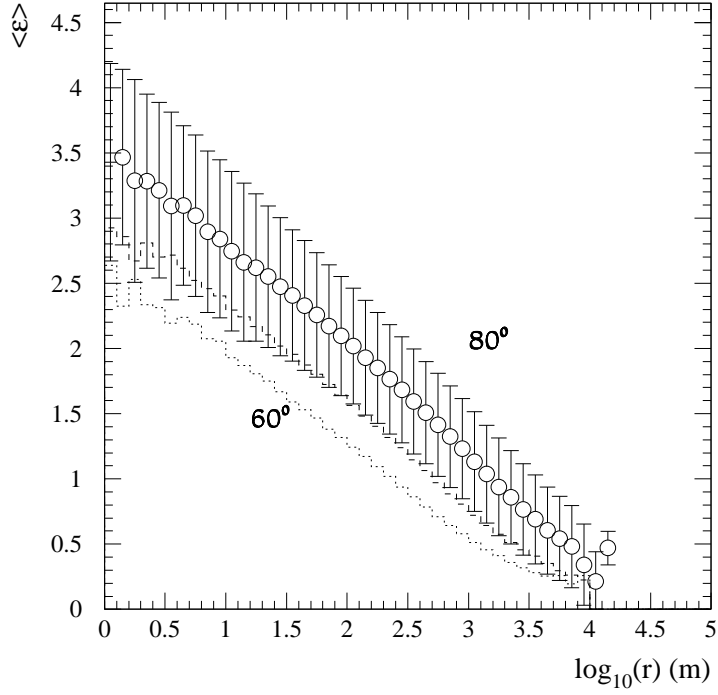,width=11.0cm}}
\caption{Correlation between $<\epsilon>=<\log_{10} E_\mu>$ and 
$\log_{10} r$ for zenith angles 60$^\circ$, 70$^\circ$, 
and 80$^\circ$ from bottom to top, see text. Error bars show the 
width of the $\epsilon$ distribution for 80$^\circ$.}
\label{epsilonfit}
\end{figure}

The quantitative model proposed still makes use of Eq.~\ref{rE}, 
keeping $d$ constant but assuming that at a given $\bar r$ there is an 
energy distribution. 
It is more convenient to work with the variable $\epsilon = \log_{10} E$. 
In this variable the correlation can be approximated: 
\begin{equation}
<\epsilon> = A - \gamma \, \log_{10} \bar r.
\label{rE083} 
\end{equation}
For all zenith angles above $60^\circ$ $\gamma$ ranges between 0.73 and 
0.83 (see table~\ref{edfparam} and Fig.~\ref{epsilonfit}). Simulations 
show that as the transverse distance to the shower axis ($\bar r$) 
increases the width of the $\epsilon$ distribution at fixed $\bar r$ 
varies slowly as shown in Fig.~\ref{epsilonfit}, being 
typically $\sim 0.4$ at $\bar r=1$~km.  

We now consider a muon ''spectral density'' in the new variable 
$\epsilon$ which is taken to be:
\begin{equation}
\rho(\bar r,\epsilon)= P(\epsilon;<\epsilon>,\sigma) ~ \rho(\bar r),
\label{maximito} 
\end{equation}
where $P$ is a distribution of mean $<\epsilon>$ and standard 
deviation $\sigma$. 
The distribution is normalized to 1 so that $\rho(\bar r)$ is recovered 
after integration in $\epsilon$. 

To obtain the muon number density in the presence of the 
magnetic field we now have to perform the same transformation:
\begin{eqnarray}
x&=&\bar x+\frac{e B_{\perp} d^2 c}{2 E}, \nonumber \\
y&=&\bar y.
\label{coord}
\end{eqnarray}
Note that this only differs subtly from Eq.~\ref{transform} as 
three variables are now involved, namely $x$, $y$ and $\epsilon$, and 
the energy $E$ in the first equation is now {\sl independent} of 
$\bar r$. The analogous expression to 
Eq.~\ref{rhotoy} for the muon number density becomes: 
\begin{equation}
\rho(x,y)=\int d \epsilon~ P(\epsilon,<\epsilon>,\sigma)~ 
\rho(\bar r),
\label{result}
\end{equation}
where in this case $\bar r$ is given by:
\begin{equation}
\bar r = \sqrt{(x-\frac{e B_{\perp} d^2 c}{2E})^2+y^2}.
\label{rbar}
\end{equation}

The advantage lies in the fact that effects ignored in the toy 
model, such as the pion transverse momentum, are taken into 
account through the energy distribution. 
Once an approximate but realistic $\epsilon$ distribution is considered 
at a fixed $\bar r$, the model transforms into a quite accurate 
parameterization of the muon number density profiles in the transverse 
plane. This is not too difficult to understand if we 
consider the density contours as described in the previous section. 
For large values of $\alpha$, which imply a shadow effect, the geometrical 
construction of the previous section implies 
largely enhanced densities accumulated at the border line 
of the shadow regions. This enhancement corresponds to the 
shock front in our Cherenkov analogy. 
The sharp edge of the distribution is an artifact of the toy model approach 
which becomes blurred when the $E$ distribution is taken into account. 
The shadow angles are radically altered, so that they are no longer 
easily related to the parameter $\alpha$. Muon number densities are also 
significantly modified with respect to the toy model in the previous section. 

\begin{figure}[hbt]
\centering
\mbox{\epsfig{figure=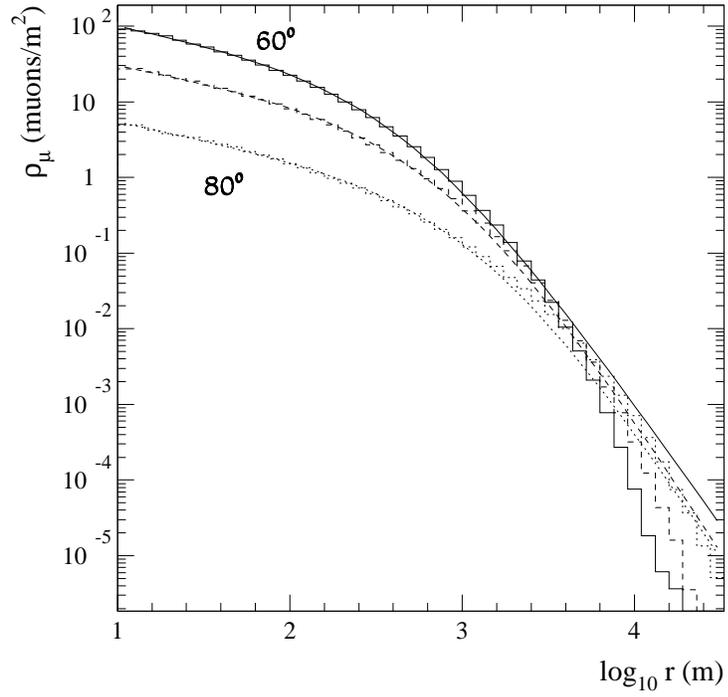,width=11.0cm}}
\caption{Lateral distribution of muons as obtained in 
the simulation and the fit given by Eq. \ref{NKGparam}. Figures 
correspond from top to bottom to zenith angles 60$^\circ$, 
70$^\circ$, and 80$^\circ$.}
\label{rhofit}
\end{figure}

Eq~(\ref{result}) requires just four inputs namely a distribution for 
$\epsilon$, the effective distance to the production 
site $d$, the relation between the mean of the distribution and 
$\bar r$ and finally the lateral distribution function of the muons in 
the transverse plane. 
We have obtained reasonable choices for the inputs from 
the results of 100 proton showers of energy $10^{19}~$eV generated 
with AIRES for each zenith angle. All of these simulations must be made in 
the absence of magnetic fields for this purpose. 
For the $\epsilon$ distribution we choose a Gaussian with mean given by 
Eq.~\ref{rE083} and a fixed intermediate value of 
$\sigma \sim 0.4$ which is accurate enough for our purposes. 
In table \ref{alturas} the production distance 
$d$ is shown for zenith angles $60^\circ$, $70^\circ$, $80^\circ$ 
and $87^\circ$. The required parameters for the mean in Eq.~\ref{rE083} 
can be read in Table~\ref{edfparam}. 
Lastly for the lateral distribution function we fit a function of the 
NKG type \cite{NKG}: 
\begin{equation} 
\rho(\bar r)=N~\bar r^{p_1} \left[ 1 + {\bar r \over a} \right] ^{p_2},
\label{NKGparam}
\end{equation}
to the simulated results. 
The results of the fits are shown in Fig.~\ref{rhofit} and the corresponding 
parameters are included in Table~\ref{ldfparam}. To take energy 
loss into account in a simple fashion we can replace $E$ at the 
transverse plane in Eq.~\ref{rbar} by the average energy along the 
muon path which is considered to start at distance $d$. The number 
densities can now be easily obtained with Eq.~\ref{result} what completes 
our model. 

\section{Ground Plane Effects}
\label{asymmetries}

Our results have been designed for the transverse plane to retain 
the maximum symmetries of the problem. 
In the event of equal number of $\mu^+$ and $\mu^-$ it is easy to see that 
muon distributions must have a very accurate mirror symmetry about 
the line defined by the transverse component of the magnetic field (see Fig.~\ref{symmetry}). 
Unfortunately a projection (extrapolation) from the transverse 
onto the ground plane or vice versa must be made to compare our results 
both with real data or with simulation. 
If we just make the obvious rectangular projection which only 
involves the cosine of the zenith angle, the results on the ground will keep 
the mirror symmetry. 

\begin{figure}[hbt]
\centering
\mbox{\epsfig{figure=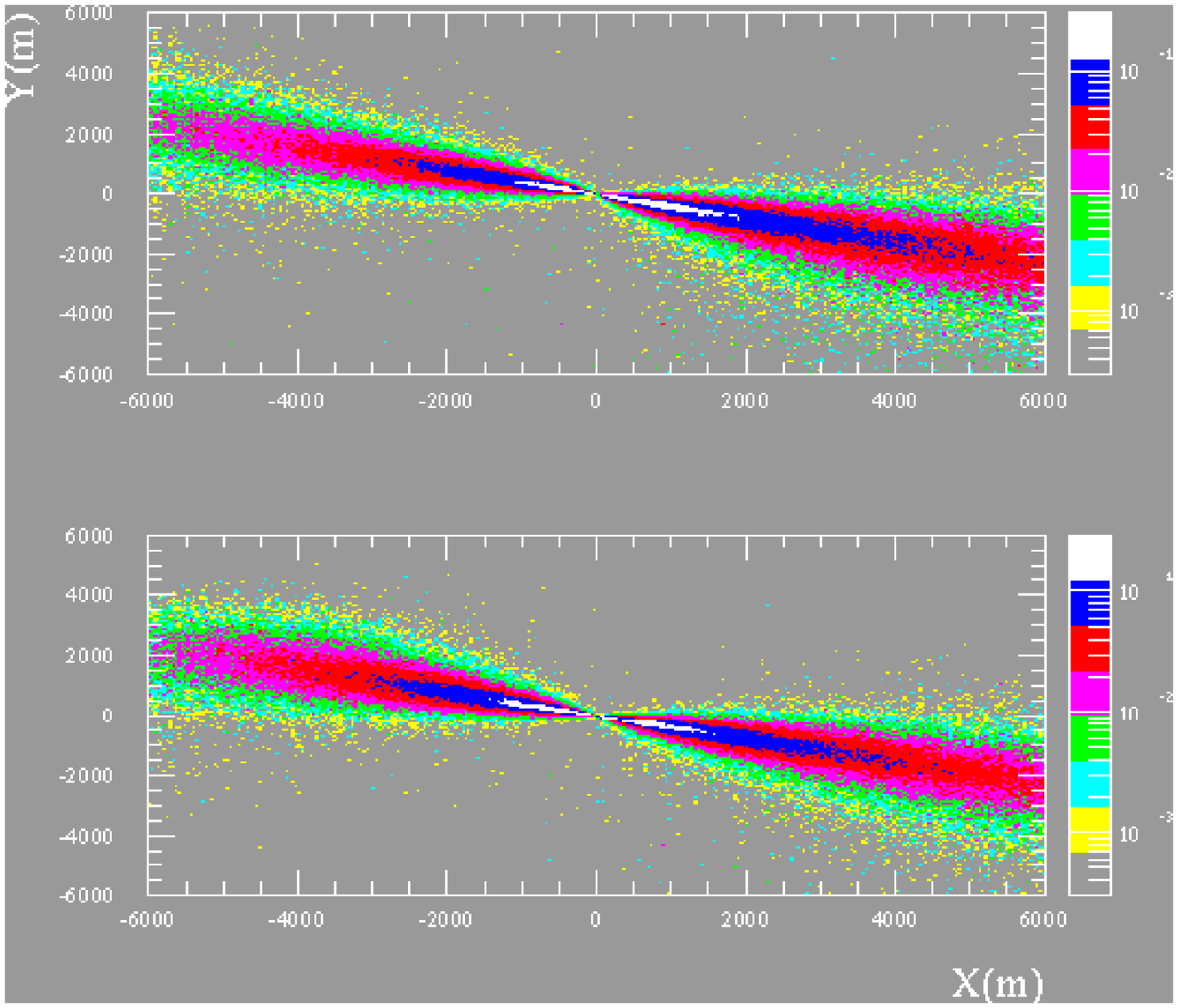,width=11.0cm}}
\caption{Muon number density contour plots  for $\phi=90^\circ$ 
and $\theta = 87^\circ$ using rectangular projection and bulk 
geometrical correction as described in the text.}
\label{asymcontour}
\end{figure}

\begin{figure}[hbt]
\centering
\mbox{\epsfig{figure=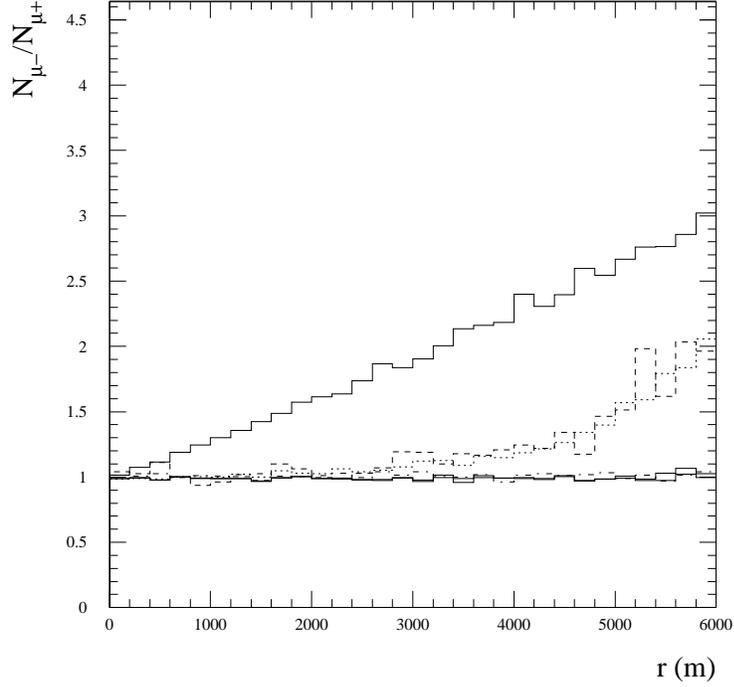,width=11.0cm}}
\caption{$\mu^+$ $\mu^-$ ratio as a function of distance to the core 
$r$ in the transverse plane for $\phi=0^\circ$ (bottom lines) and for 
$\phi=90^\circ$ (upper lines) using rectangular projection 
(continuous lines), individual muon extrapolation (dashed lines), 
and bulk geometrical correction (dotted lines), as described in the 
text.}
\label{mupluminus}
\end{figure}

At high zenith angles however the results of a simulation 
show some distortion of this mirror symmetry, see Fig.~\ref{asymcontour}. 
Clearly half of transverse plane is 
''below'' ground level which means that in a real situation the muons 
arrive to 
the ground before reaching the transverse plane while the other half is 
''above'' so that the muons have to travel more atmosphere before hitting 
the surface of the Earth. One way of quantifying the asymmetry is through 
$\mu^+$ $\mu^-$ ratio shown in Fig.~\ref{mupluminus}. 
The $\mu^+$ and $\mu^-$   
are completely separated into two lobes which in general arrive at 
different times to the ground. As expected 
this effect is largest when the angle $\psi$ (see Fig.~\ref{geometry}) is 
smallest (corresponding to azimuth $\phi=90^\circ$) 
but disappears at $\psi=90^\circ$ (when the shower is aligned 
with the magnetic North $\phi=0^\circ$).  

The asymmetry arises because of the different muon paths from the 
production point to ground level. The cumulative effects 
of all muon interactions (such as magnetic deflections and continuous 
energy loss) and decays depend on the muon path. As a result the 
asymmetries have a markedly geometrical nature 
growing as the distance to shower core increases. 
This is fortunate since simple 
geometrical algorithms can be devised to correct these asymmetries 
when necessary. 

The two sources of muon deviations from shower axis that have been 
considered in previous sections account for most of the geometric 
effect. For each zenith angle the muon paths are considered to 
start at a fixed distance $d$ to the transverse plane. Corrected 
projections must take this common origin of the muon paths into account 
that clearly generates an asymmetry on the ground plane. 
The muon deviations due to the magnetic field are also a clear source 
of asymmetry, which is quadratic on the muon path (arcs of circle).  
When this is taken into account the simulated results can 
be ''projected'' onto the transverse plane in a way that most of the 
asymmetries arising at high zenith angles can be removed. 

There are several possibilities available for projection, 
the easiest being the obvious rectangular projection. 
Such a projection is enough for most purposes because the asymmetries 
are small, except for extreme 
zenith angles. For such cases the ''quadratic'' deviation due to 
the magnetic field becomes most important and we have two ways 
to account for it depending on whether we work with simulated results or not. 
We can project simulated results onto the transverse plane just 
extrapolating the individual muon trajectories since all the information on 
the particles reaching the ground is easily available in a simulation. 
Alternatively one can use approximate geometrical projections which will 
be shown to be sufficiently accurate in spite of ignoring 
the details of the individual muons. This is because of the discussed correlation between energy and position of the muons.  

Clearly individual muon extrapolation is the best way of obtaining the 
muon densities in the transverse plane. A simple helix would 
only neglect muon energy loss and decay and some correction can be 
further implemented to account them. 
For a purely geometrical extrapolation we will use the following procedure 
which is shown to be sufficient for all purposes. 
In the direction parallel to 
$B_{\perp}$ one simple joins the muon position to the fixed production 
point while in the orthogonal direction one considers muon trajectories 
which are circles tangent to shower axis at the same production point. 
In Figs.~\ref{asymcontour} and in Fig.~\ref{mupluminus} the results of 
the rectangular projection are compared to the extrapolation 
of the simulated densities for a shower at extreme zenith 
$\theta=87^\circ$ and having azimuth $\phi =90^\circ$ (maximum asymmetry). 
A large fraction of the unwanted asymmetries in the transverse plane 
is equivalently removed with either of the two methods. 
The residual asymmetries are mostly due 
to muon decay and the neglect of muon energy loss. 
For distances to shower 
core below $4$~km the 
residual asymmetry is below $20 \%$; it should be stressed that this range corresponds to distances up to 80~km away from shower core in the ground 
plane.  

In the analytical results obtained for the muon densities in the 
transverse plane (previous section) the information on the individual 
muon energy and direction is lost. 
Fortunately the geometrical algorithm to remove the asymmetries can be 
inverted to convert the symmetric results obtained in our model to the 
ground, approximately reproducing the asymmetries, what makes it 
appropriate for comparison with data to the precision level indicated 
by these graphs. The systematic error in the ground projection 
that would be generated at large $r$ could be largely removed 
accounting for asymmetries due to energy loss and muon decay. 

\section{Results}
\label{results}

Testing the muon number density results obtained with the 
prescription described in the previous section by comparison with 
dedicated simulations is by no means a straightforward task. The 
main difficulty arises from the uncertainties associated with the 
muon densities on the ground plane that are obtained with any 
simulation program which necessarily implies the statistical 
treatment of shower particles, {\sl thinning algorithms}, 
for the energies considered to avoid excessive computing times. 
There is a complex combination of statistical fluctuations, shower 
fluctuations and thinning fluctuations which is rather difficult to 
unravel. The simulated muon number density results fluctuate 
depending on the two dimensional binning choices which are rather 
awkward to choose because of the complexity of the patterns 
involved particularly at the highest zenith angles. 

For testing purposes, we choose to compare the densities obtained 
with the described prescription to the results of simulations using 
AIRES \cite{Aires} with the magnetic effects on. We have also 
generated 100 proton showers of energy $10^{19}~$eV and azimuths 
$\phi=0^\circ$ and $90^\circ$ for each zenith angle ($60^\circ$, 
$70^\circ$, $80^\circ$, and $87^\circ$). We project the average muon 
densities (over the 100 showers) onto the transverse plane as 
described in the previous section to eliminate the asymmetries and 
to compare to the results of our model. 

\begin{figure}[hbt]
\centering
\mbox{\epsfig{figure=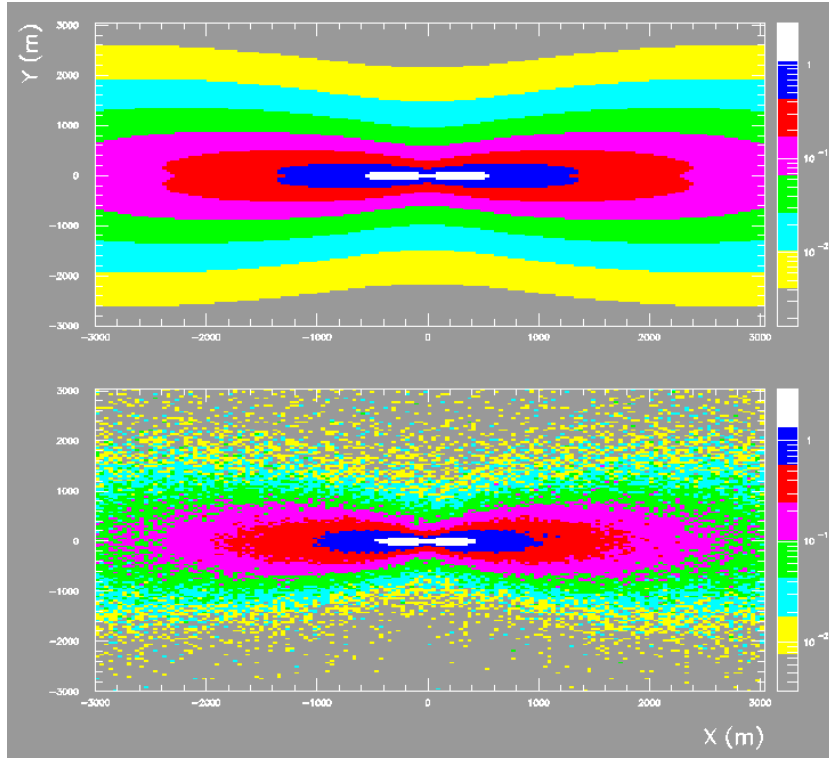,width=11.0cm}}
\caption{Contour plots of the muon density in the transverse 
plane for $10^{19}$~eV proton showers with an incident zenith angle 
of 80$^\circ$ as obtained in the simulation for $\phi=0^{\circ}$ 
(lower panel) and in the model (upper panel).}
\label{contour80}
\end{figure}

\begin{figure}[hbt]
\centering
\mbox{\epsfig{figure=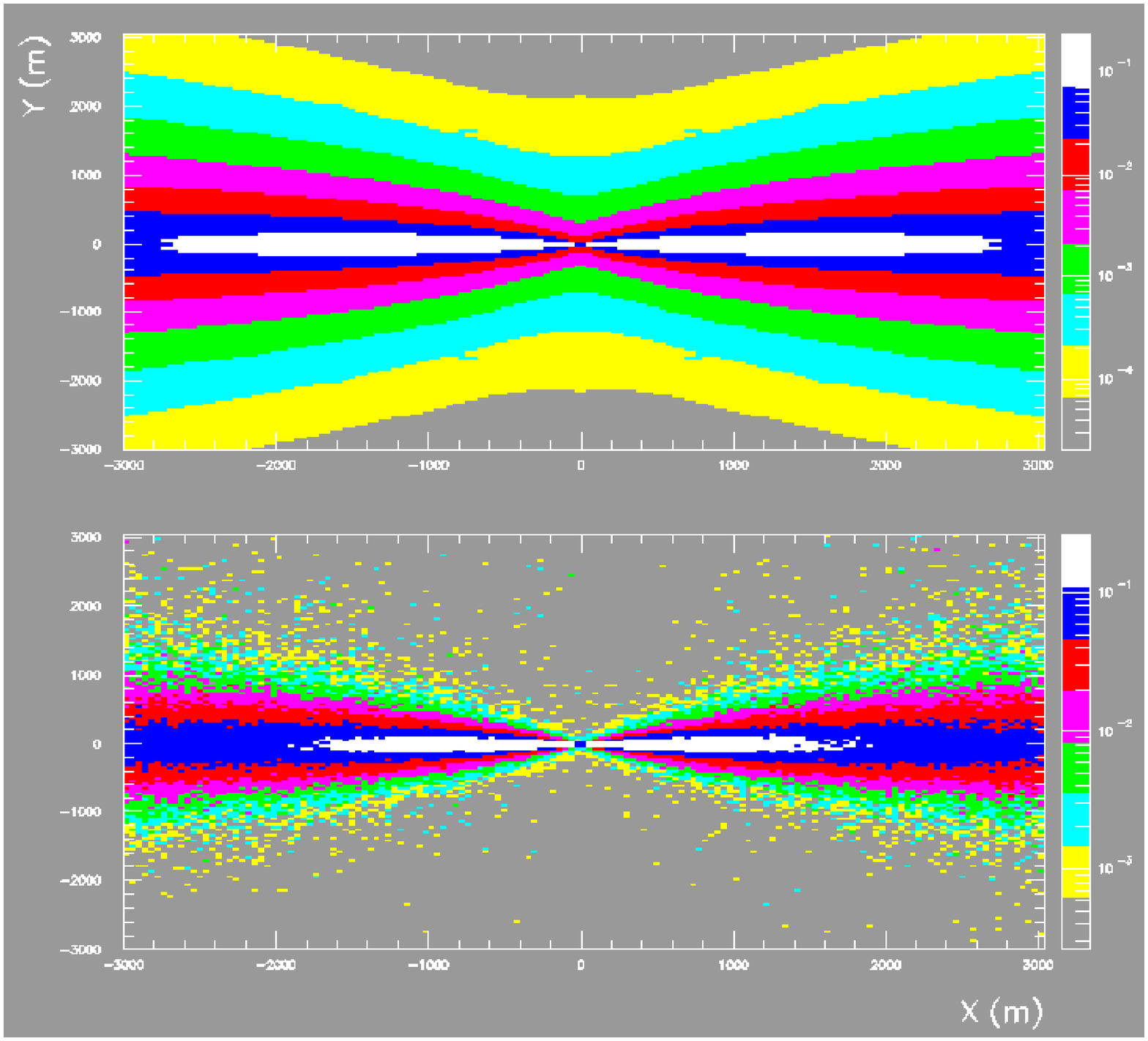,width=11.0cm}}
\caption{Same as figure \ref{contour80} but for incident zenith angle of
$87^\circ$.}
\label{contour87}
\end{figure}

\begin{figure}[hbt]
\centering
\mbox{\epsfig{figure=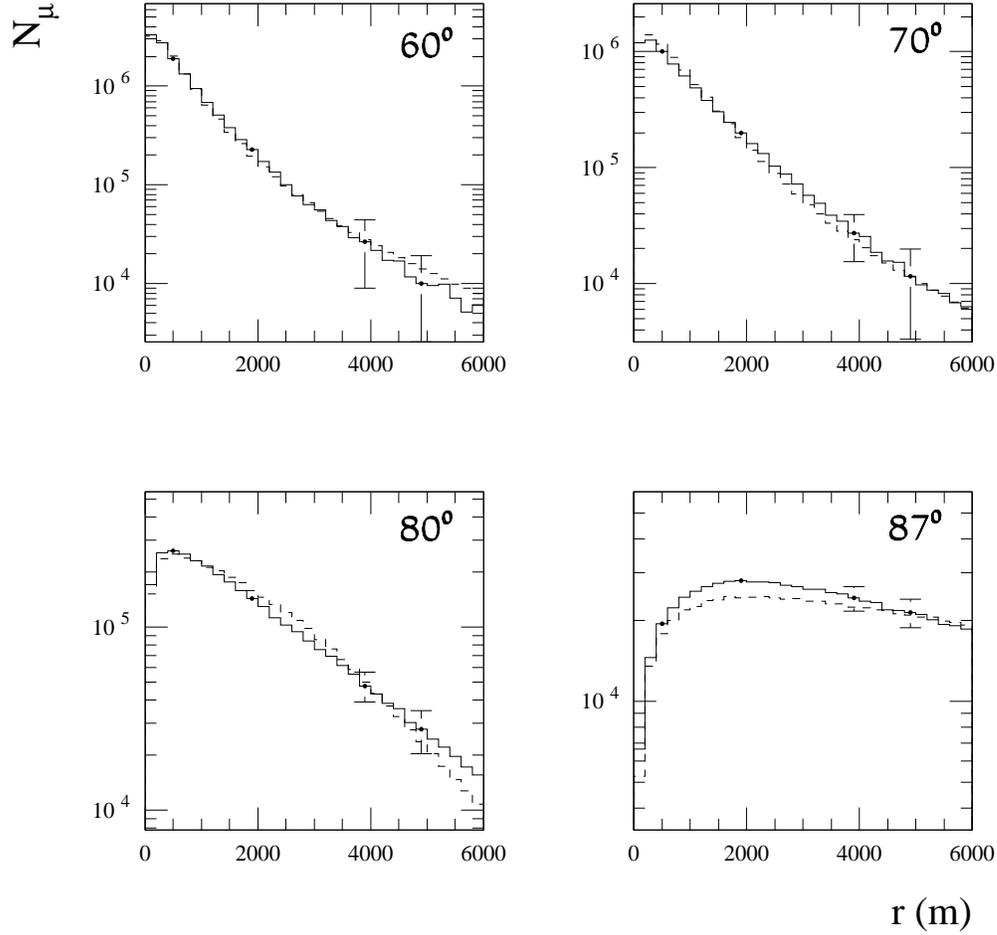,width=15.0cm}}
\caption{Transverse distance ($r$) distributions of the muons 
in the transverse plane obtained with $10^{19}$~eV proton shower 
simulations (continuous histograms) compared to the model (dashed lines) 
for zenith angles 60$^\circ$, 70$^\circ$, 80$^\circ$, 87$^\circ$. The 
error bars shown include both the statistical error and the shower 
to shower fluctuations.}
\label{phiint}
\end{figure}

\begin{figure}[hbt]
\centering
\mbox{\epsfig{figure=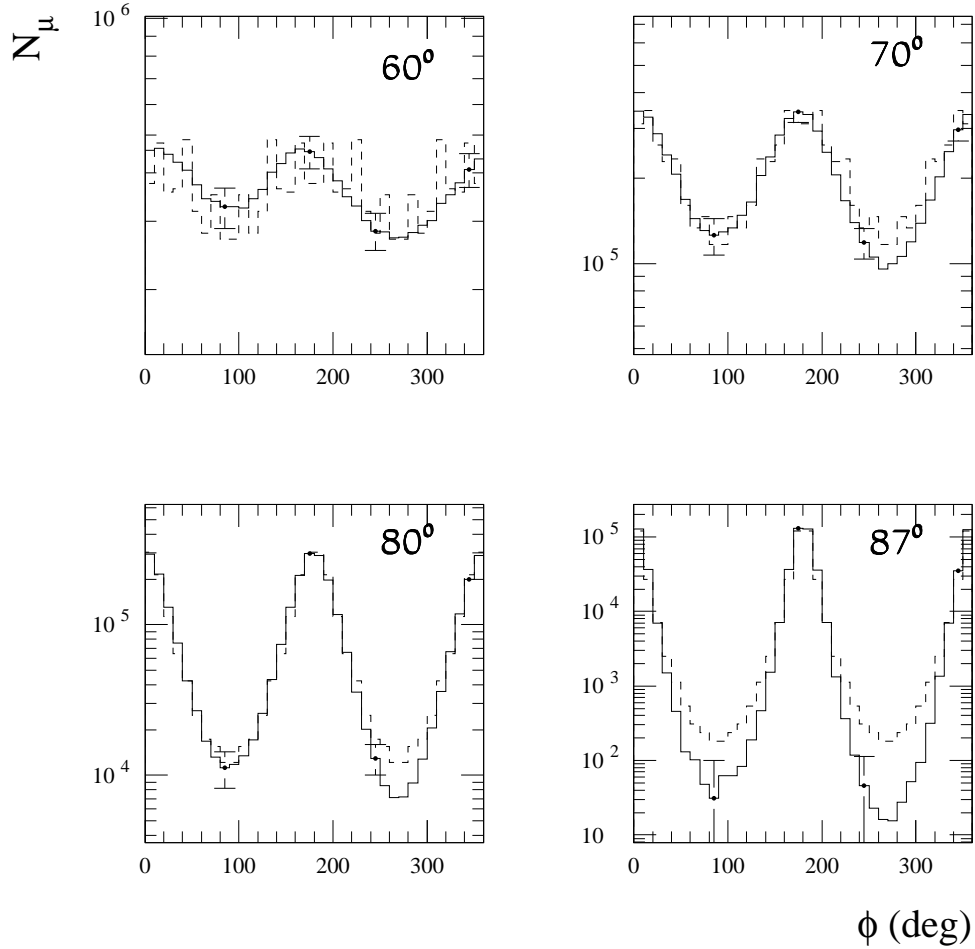,width=15.0cm}}
\caption{Azimuthal angle distribution of muons in the transverse 
plane (integrated up to $r=6$~km) as obtained in 
$10^{19}$~eV proton shower simulations (continuous histogram) 
compared to the model (dashed histogram) for zenith angles 
60$^\circ$, 70$^\circ$, 80$^\circ$, and 87$^\circ$.}
\label{rint}
\end{figure}

\begin{figure}[hbt]
\centering
\mbox{\epsfig{figure=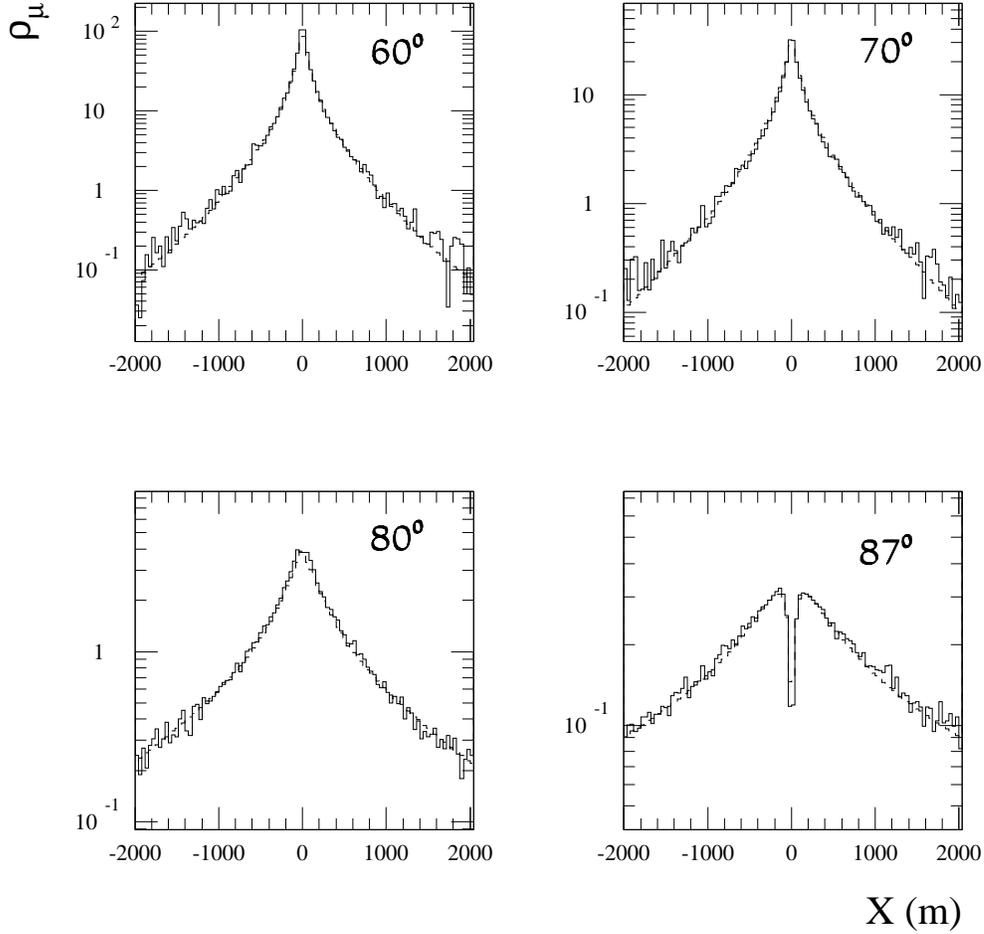,width=15.0cm}}
\caption{Muon number density in the transverse plane as a function of 
the $x$ coordinate for a fixed value of $y$ as obtained in 
$10^{19}$~eV proton showers simulations (continuous lines) for 
$\phi=0^{\circ}$ compared 
to the model (dashed lines). Densities are calculated in $40 \times 40$~m 
bins and the $x$ bins shown are centered at $y=0$. 
The figures correspond to zenith angles 
60$^\circ$, 70$^\circ$, 80$^\circ$, and 87$^\circ$.}
\label{xslices}
\end{figure}

\begin{figure}[hbt]
\centering
\mbox{\epsfig{figure=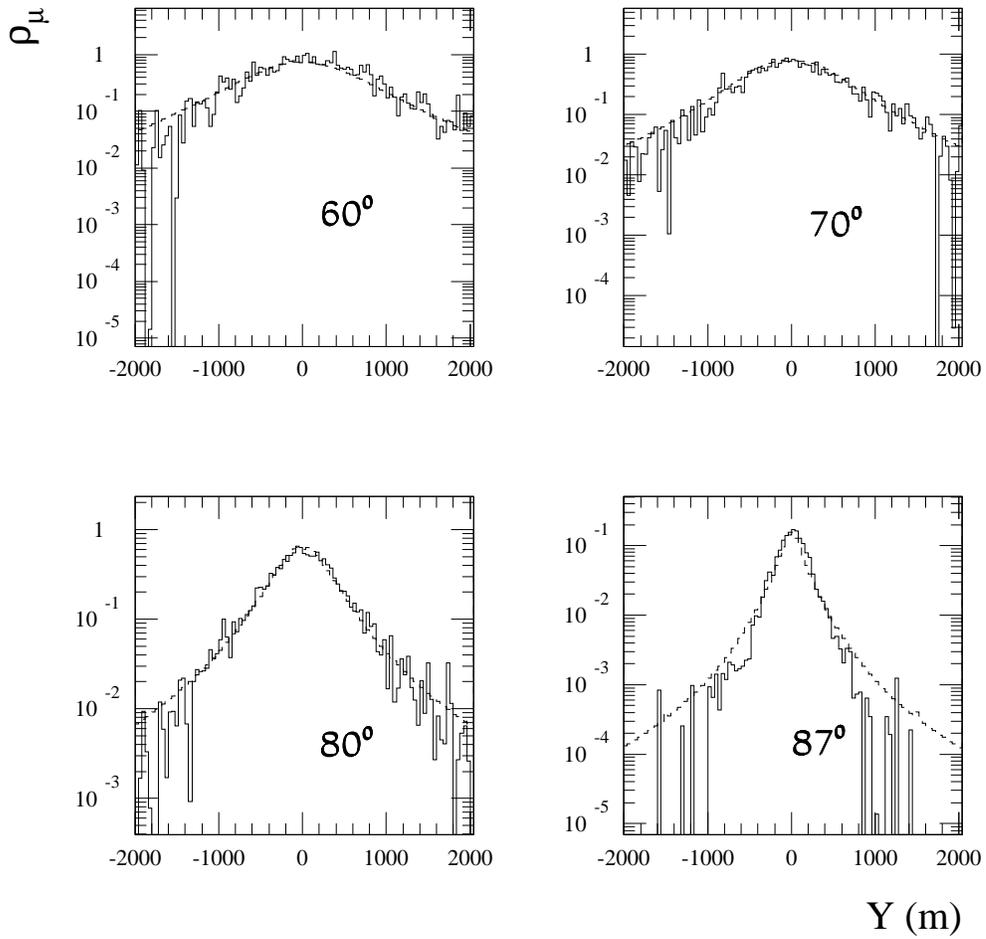,width=15.0cm}}
\caption{Same as figure \ref{xslices} but plotting densities as a function 
of $y$ at fixed $x$. Densities shown correspond to $y$ bins centered at 
$x=1000$~m.}
\label{yslices}
\end{figure}

The number density results of Eq.~(\ref{result}) are compared to the 
averages obtained with the specific simulations made. No normalization 
has been made in the comparisons shown in 
Figs.~\ref{contour80}--\ref{yslices} that illustrate the reach of our 
results. The figures shown very much speak on their own. 
The agreement is rather impressive considering the ranges of 
distances involved and particularly the ranges in densities which can span 
over four orders of magnitude. 
We firstly show in Figs.~\ref{contour80} and \ref{contour87} 
two density contour plots for two different zeniths. 
These graphs display well defined shadow regions as anticipated and 
show an excellent agreement in the shape of the densities obtained 
compared to the simulation results. The isolated 
density spots in the simulation also indicate very 
clearly how the fluctuations due to the simulation become 
very important even at distances very close to shower axis in some 
specific directions. The figures even suggest that in many respects 
the parameterization results are ''better behaved'' in the sense that 
they avoid many unphysical fluctuations and can be extrapolated to 
very low muon densities. 

Quantitative tests of the average two dimensional density distribution 
in polar coordinates integrated over the radial 
(azimuthal) coordinate up to 6~km (up to $2\pi$) in the transverse 
plane are shown in Fig.~\ref{phiint} (Fig.~\ref{rint}). The 
bars are indicative of the statistical errors in the obtained averages  
taking into account shower fluctuations and poisson statistics. 
These graphs 
are particularly good for stressing systematic effects that are 
cumulative and highlight the discrepancies. The disagreement is 
largest for the highest values of $r$ so that the highest 
discrepancies correspond to low muon number densities where results 
are going to be less relevant. As stressed before some of the differences 
between the model and the simulation are due to muon decay and are thus 
still an artifact of using results on the ground 
plane for comparison with our analytical muon number densities. They 
are nevertheless indicative of the uncertainties that can 
be expected if geometrical projections are used for obtaining densities 
at ground level ignoring muon decay. These are below statistical 
fluctuations and consequently quite hidden by shower fluctuations. 

Lastly, Fig.~\ref{xslices} shows the $x$ distribution calculated in a 
40~meter bin centered at $y=0$ while Fig.~\ref{yslices} shows the $y$ 
distribution for a 40~meter bin centered at $x=1~$km. 
Although for the highest zenith ($87^\circ$) the $r$ 
distributions disagree above the $20\%$ level for 
$r$ distances exceeding 4~km in the transverse plane, the discrepancies 
in the actual densities correspond to very low values of the densities 
which are relatively less important. 

\section{Summary and Discussion}
\label{summary}

We have successfully obtained a framework for studying the muon content 
of HAS produced by protons in the atmosphere. It can be 
effectively considered as a semianalytical parameterization for the 
muon number densities in the transverse plane for HAS. 
Although 
it relies on Monte Carlo it can be considered as an alternative to it 
since the simulation is only needed in first instance to obtain the inputs 
ignoring the magnetic effects and hence eliminating azimuthal angle 
dependences. Although only proton showers of $10^{19}~$eV energy have 
been considered for the Haverah Park site, this framework is completely 
general and can be applied to showers of any energy and composition 
located in any site in a completely analogous fashion. 

The method 
consists of splitting the muon propagation in a shower into two fictitious 
parts, one which takes place in the absence of a magnetic field and 
a second one in which the magnetic effects are separately analysed. 
As the first part preserves the approximate circular symmetry of the 
shower development it allows simple parameterizations in terms of a 
single variable. The second part can be implemented in an analytical 
way. 
As a spinoff we have obtained a way to obtain ground density profiles 
of fixed zenith angle but any azimuthal angle from a single shower of the 
corresponding zenith. 

We expect to refine this model by improving the fits 
for the required inputs and including ignored effects such as 
correlations between muon impact position and distance to production site, 
muon decay, and distributions of distance to 
production site. These ignored effects could be implemented in an 
analogous way to the introduction of the energy distribution. 
Although these refinements may be necessary in the future, we do not 
feel at this time there is need to be so accurate since there are 
many uncertainties in the simulation programs which surely exceed 
the precision level of the results obtained. The simulations used 
require extrapolations of cross sections and distributions of 
particle interactions in energy 
regions which are well above those explored in current particle physics 
experiments or those to be made in the near future. 

Besides giving insight into the complexity of large zenith angle 
showers, the results presented here have a rather broad range of 
applications connected with the detection of Horizontal Air Showers. 
The need of systematic simulations of complicated high energy 
showers which is very time consuming can be avoided. Such a task 
is of great help for the analysis of large zenith angle data as well 
as for the study of the sensitivity of these air showers to any of 
the current concerns about the highest energy cosmic rays such 
as the expected acceptance of Pierre Auger Observatories \cite{Auger}.

{\bf Acknowledgements:} 
We are indebted to A.A.~Watson for having introduced us into the 
real experimental problems of HAS and an invitation to Leeds when 
most of this work was initially conceived. We also thank J.A.~Hinton, 
G.~Parente, and A.A.~Watson for many discussions and also for 
comments after reading our manuscript.  
This work was partly supported by a joint grant from the 
British Council and the spanish Ministry of Education (HB1997-0175), 
by Xunta de Galicia (XUGA-20604A98) and by CICYT (AEN99-0589-C02-02). 
%
  
%

%
%
\begin{table}
\begin{displaymath}
\begin{array}{||c|c|c|c|c|c|c||}
\hline
\hline
\theta \mbox{(degrees)} & \mbox{d (km)} & \mbox{$\Delta$d (km)} & 
\mbox{ $<E>$ (GeV)} & \mbox{ $\Delta E$ (GeV)} & 
N_\mu \times 10^{-6}& \Delta N_\mu  \times 10^{-6} \\
\hline
0^\circ  & 3.9   & 2.8 & 8.1  & 33.3 & 29.  & 6.5  \\
60^\circ & 16    & 6.5 & 18.9 & 90   & 13.3 & 2.4  \\
70^\circ & 32    & 10  & 32.9 & 208  & 7.8  & 1.4  \\
80^\circ & 88    & 17  & 77   & 486  & 3.3  & 0.7  \\
87^\circ & 276   & 31  & 204  & 1186 & 1.2  & 0.2  \\
\hline
\hline
\end{array}
\end{displaymath}
\caption{Relevant parameters for muon production as obtained in 100 
proton showers of energy $10^{19}$ eV with a relative thinning of
$10^{-6}$ simulated with AIRES using SIBYLL 1.6 cross sections. 
Average values and RMS deviations for production altitude (d), 
muon energy at production ($<E>$), and total number of muons at 
ground level ($N_\mu$).}
\label{alturas}
\end{table}

\begin{table}
\begin{displaymath}
\begin{array}{||c|c|c||}
\hline
\hline
\theta \mbox{(degrees)} & \gamma & A \\
\hline
60^\circ & 0.75 \pm 0.1  & 2.67 \pm 0.23 \\
70^\circ & 0.73 \pm 0.06 & 4.04 \pm 0.20 \\
80^\circ & 0.81 \pm 0.07 & 3.63 \pm 0.20 \\
87^\circ & 0.83 \pm 0.07 & 4.15 \pm 0.20 \\
\hline
\hline
\end{array}
\end{displaymath}
\caption{Best values for the parameters in the relation 
between average muon energy and distance to shower core (Eq.~8) 
and corresponding errors, as obtained in the fit to the simulated 
results.}
\label{edfparam}
\end{table}

\begin{table}
\begin{displaymath}
\begin{array}{||c|c|c|c|c||}
\hline
\hline
\theta \mbox{(degrees)} & N & p_1 & p_2 & a \mbox{(m)} \\
\hline
60^\circ & 569.9 & -0.52 & -4.05 & 782.8 \\
70^\circ & 227.1 & -0.49 & -4.35 & 1010  \\
80^\circ &  78.4 & -0.52 & -4.49 & 1513  \\
87^\circ &  20.2 & -0.51 & -5.43 & 2171  \\
\hline
\hline
\end{array}
\end{displaymath}
\caption{Best values for the parameters in the lateral distribution 
function in Eq.~13 as obtained in a fit to the simulated results.} 
\label{ldfparam}
\end{table}

%
%
%

%

\end{document}